\documentclass[12pt,article]{JHEP3}
\usepackage{amsmath,amssymb,euscript,array,mathrsfs,epsfig}
\newcommand{\startappendix}{
\setcounter{section}{0}
\renewcommand{\thesection}{\Alph{section}}}
\newcommand{\Appendix}[1]{
\refstepcounter{section}
\begin{flushleft}
{\large\bf Appendix \thesection: #1}
\end{flushleft}}
\def\ben
{\begin{equation}}
\def\een{\end{equation}}

    \let\L=\Lambda
   
 \let\W=\mu

\def\W={\cal W}
\def\L ={\cal L}

\def\be{\begin{equation}}
\def\ee{\end{equation}}
\def\ba{\begin{array}}
\def\ea{\end{array}}

\def\dalemb#1#2{{\vbox{\hrule height .#2pt
        \hbox{\vrule width.#2pt height#1pt \kern#1pt
                \vrule width.#2pt}
        \hrule height.#2pt}}}

\newcommand{\bea}{\begin{eqnarray}}
\newcommand{\eea}{\end{eqnarray}}

\title{Thermodynamics of higher spin black holes in 3D}
\author{Justin R. David${}^{a}$, Michael Ferlaino${}^{b}$ and 
S. Prem Kumar${}^{b}$\\\\
${}^{a}${\it 
Centre for High Energy Physics,\\
\,\,Indian Institute of Science, \\ 
\,\,C. V. Raman Avenue, Bangalore 560012, India.  
}\\
\,\,\,\email{justin@cts.iisc.ernet.in}
\\\\
${}^{b}${\it 
Department of Physics,\\\,\,\,Swansea University, \\ 
\,\,\,Singleton Park, Swansea, SA2 8PP, U.K.  
}\\
\,\,\,\email{pymf@swansea.ac.uk, s.p.kumar@swansea.ac.uk}
} 
\abstract{We examine the thermodynamic properties of recently constructed black hole solutions in ${\rm SL}(3,\mathbb R) \times {\rm SL}(3,\mathbb R)$ Chern-Simons theory in the presence of a chemical potential for spin-3 charge, which acts as an irrelevant deformation of the dual CFT with ${\cal W}_3\times {\cal W}_3$ symmetry. The smoothness or holonomy conditions admit four branches of solutions describing a flow between two  ${\rm AdS}_3$ backgrounds corresponding to two different CFTs. The dominant branch at low temperatures, connected to the BTZ black hole, merges smoothly with a thermodynamically unstable branch and disappears at higher temperatures. 
We confirm that the UV region of the flow satisfies the Ward identities of a CFT with ${\cal W}_3^{(2)}\times {\cal W}_3^{(2)}$ symmetry deformed by a spin-$\tfrac{3}{2}$ current. This allows to identify the precise map between UV and IR thermodynamic variables.
We find that the high temperature regime is dominated by a black hole branch whose thermodynamics can only be consistently inferred with reference to this ${\cal W}_3^{(2)}\times {\cal W}_3^{(2)}$ CFT.
}
\begin{document}
\section{Introduction and Summary}
The correspondence between conformal field theories in $d$-dimensions and gravitational theories on anti-de-Sitter (AdS) spacetimes in $d+1$-dimensions \cite{maldacena,witten,magoo}, provides a natural setting and motivation for the study of higher spin theories of gravity \cite{Fronsdal:1978rb, Fradkin:1987ks, Vasiliev:1990en}. In weakly interacting limits large-$N$ field theories have been argued to be dual to higher spin theories of gravity in AdS spacetimes \cite{Konstein:2000bi, HaggiMani:2000ru, Sundborg:2000wp, Mikhailov:2002bp, Sezgin:2002rt}. 
Perhaps most notable in this context is the Klebanov-Polyakov proposal relating the $O(N)$ vector model at large $N$ in three dimensions \cite {Klebanov:2002ja, Giombi:2009wh,Giombi:2010vg} to Vasiliev's higher spin theory on ${\rm AdS}_4$. 

The complexity of higher spin theories reduces drastically in three dimensions wherein it becomes possible to consistently truncate to a finite set of higher spin fields \cite{blencowe} and reformulate the theory with spin $s\leq N$ in terms of 
${\rm SL}(N,\mathbb R) \times {\rm SL}(N,\mathbb R)$ Chern-Simons theory 
\cite{Bergshoeff:1989ns,Bordemann:1989zi,Vasiliev:1989qh,Vasiliev:1989re,Fradkin:1990qk,Henneaux:2010xg,campoleoni}. The works of \cite{Henneaux:2010xg,campoleoni} have shown that the asymptotic symmetry algebra 
of such higher spin theories on ${\rm AdS}_3$ is (two copies of) the ${\cal W}_{N}$ algebra. This naturally generalizes the formulation of ordinary gravity on ${\rm AdS}_3$ in terms of ${\rm SL}(2,\mathbb R) \times {\rm SL}(2,\mathbb R)$ Chern-Simons theory \cite{Achucarro:1987vz, Witten:1988hc} and extends the associated asymptotic Virasoro symmetry \cite{Brown:1986nw} to the ${\cal W}_N$ algebra. Spurred on by these developments another concrete proposal by Gaberdiel and Gopakumar  \cite{gg} posits a duality between three dimensional higher spin theories and a ${\cal W}_N$ minimal model CFT in a 't Hooft-like large-$N$ limit \cite{Gaberdiel:2011zw, Gaberdiel:2011wb, Chang:2011mz, Ahn:2011pv}. 

Given that higher spin theories are non-trivial extensions of ordinary gravity, it is important to understand the nature of classical solutions in such theories which generalize the notion of diffeomorphism invariance to a higher spin gauge symmetry. Recently, Gutperle and Kraus \cite{gutkraus} provided a construction of black hole like solutions in ${\rm SL}(3,\mathbb R) \times \rm{SL}(3,\mathbb R)$ Chern-Simons theory which can be viewed as gravity coupled to a spin-3 field on ${\rm AdS}_3$. The dual 2d CFT has ${\cal W}_3\times{\cal W}_3$ symmetry generated by the stress tensor and a spin-3 current \cite{Zamolodchikov:1985wn}. Interestingly, it turns out that the black hole solutions which carry spin-3 charge are actually embedded within a renormalization group (RG) flow between two CFT's. The reason is that the current which generates the spin-3 symmetry is a dimension-3 operator in the dual CFT, and a chemical potential for the higher spin charge acts as an irrelevant deformation in the boundary CFT with ${\cal W}_3\times {\cal W}_3$ symmetry. As such, an irrelevant coupling in a quantum field theory would be potentially problematic in the ultraviolet (UV). Nonetheless, the gravity/Chern-Simons description exhibits a flow to a new UV fixed point CFT with ${\cal W}_3^{(2)}\times {\cal W}_3^{(2)}$ symmetry. The ${\cal W}_3^{(2)}$ algebra is generated by the stress tensor, two spin-$\tfrac{3}{2}$ currents and a spin-1 current, and is also known as the Polyakov-Bershadsky algebra \cite{Polyakov:1989dm, Bershadsky:1990bg}. 

In addition to having non-vanishing higher spin charge, a black hole solution potentially describing the finite temperature state of a field theory interpolating between two CFT's is also extremely interesting in its own right. Since the metric and associated geometrical invariants are not gauge-invariant objects in ${\rm SL}(3,\mathbb R)$ Chern-Simons theory, black holes are characterized in terms of the holonomy of the Chern-Simons gauge connection around the Euclidean time circle. Specifically, the requirement that this holonomy be trivial, replaces the usual condition that the Euclidean thermal circle vanish smoothly at the location of a black hole horizon. Remarkably, it was shown in \cite{gutkraus} that the holonomy conditions actually are equivalent to thermodynamic integrability conditions. These guarantee that the black hole mass and charge obtained from the holonomy conditions, viewed as functions of temperature and spin-3 chemical potential, can be consistently derived from a thermodynamical free energy (see also \cite{Kraus:2011ds} for generalization to ${\rm SL}(N,\mathbb R)$).  
\begin{figure}[h]
\begin{center}
\epsfig{file=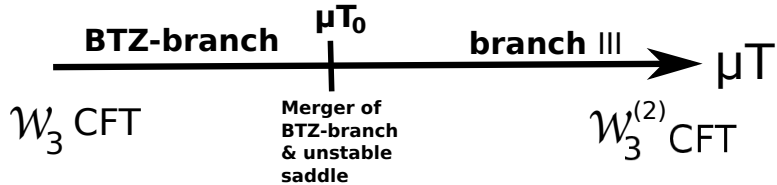, width =4.0in}
\end{center}
\caption{\small{ Two different black holes dominate the ensemble at low and high temperatures. The BTZ-branch which reduces to the ordinary BTZ black hole at $\mu T=0$, merges with an unstable saddle at $\mu T_0\,=\,\tfrac{3}{16\pi}\sqrt{2\sqrt3 -3}$ and disappears. The high temperature partition function is dominated by the stable branch-III.}}
\label{fig1} 
\end{figure}

The main aim of this paper is to examine the thermodynamic properties of all the possible solutions of the holonomy conditions. The latter are algebraic constraints on the mass and charge, and possess multiple solutions. We find that for general non-vanishing spin-3 chemical potential $\mu$ and temperature $T$, there are four possible branches of solutions of which only one, which we refer to as the ``BTZ-branch'', smoothly connects to the BTZ black hole at zero spin-3 chemical potential and which was the subject of the recent works \cite{gutkraus,ammon,Castro:2011fm,review,Kraus:2012uf}
(see \cite{Tan:2011tj} for a black hole solution with a non-vanishing spin-4 field ). 
Of the three additional branches, one (branch-II) is thermodynamically unstable (similarly to the small black hole in ${\rm AdS}_5$), the second is thermodynamically stable (branch-III), and the last one is thermodynamically disfavoured for all temperatures and chemical potentials. 

The chemical potential $\mu$ for the spin-3 charge, which has scaling dimension $-1$, and the temperature $T$ are the only dimensionful parameters in the problem. Hence the phase structure of the system depends on the dimensionless combination $\mu T$. 
 With increasing  $\mu T$ we find a rich phase diagram wherein the BTZ-branch and the unstable branch of solutions merge at a specific value of $\mu T$, and cease to exist for higher temperatures. Increasing the temperature further results in the system making a discontinuous transition to the one remaining, stable black hole solution. 
Although this appears to be a consistent picture at first sight, we find that unravelling the high temperature behaviour of thermodynamic quantities such as energy and entropy for this branch is subtle and unusual, and poses certain puzzles. 

Briefly, we find that a consistent interpretation of the energy and entropy of the high temperature solution can only be given in terms of the variables appropriate for describing the UV fixed point CFT with ${\cal W}_3^{(2)}$ symmetry. The key point is that the stress tensors of the UV and IR CFT's are not related.  The thermodynamical charges of the two fixed point CFT's undergo a relabelling along the RG flow. The limit of large $\mu T$ is most naturally interpreted as a relevant deformation of the ${\cal W}_3^{(2)}$ CFT by a chemical potential $\lambda\sim 1/\sqrt\mu$ for spin-$\tfrac{3}{2}$ currents. We find that the high temperature expansions of the free energy and entropy, when appropriately defined with respect to the UV fixed point, do have the natural dependence on temperature expected from a CFT in two dimensions. Our analysis confirms that the holonomy conditions (and their solutions) correctly capture  non-trivial aspects of the RG flow between the two CFT's in question.

 In order to infer the correct global thermodynamics, we must have a  prescription to compute the grand potential. In principle this is  encoded in the holonomy conditions which are thermodynamic integrability conditions as pointed out above. However, given the non-linearity of the resulting algebraic equations, and the possibility of multiple solutions to them, it is not {\em a priori} clear how to obtain a closed form for the grand potential which applies to all solutions. The answer to this question was recently provided in \cite{bct} (see also \cite{Perez:2012cf} ) and we apply their proposal to compute explicitly the grand potential for the system. This involves a careful evaluation of the Chern-Simons boundary action and its variation. We find an extremely simple form for the result which satisfies all consistency checks, and which we subsequently employ to discuss the thermodynamics of the different branches.

The metric description of all branches of solutions involves wormhole geometries \cite{gutkraus} along with background spin-3 fields. This is, of course, a gauge dependent description and it was demonstrated explicitly in \cite{ammon} that the wormhole geometries can be gauge transformed to the so-called black hole gauge. In this gauge the metric has a horizon where the Euclidean time circle shrinks smoothly to zero. Our analysis reveals that the black hole gauge metric of \cite{ammon} applies only to the BTZ-branch and the unstable branch-II (which merge at a certain temperature). In fact the unstable branch-II attains its maximal value of  spin-3 charge at zero temperature where it has non-vanishing entropy, and precisely matches the extremal solution discussed in \cite{ammon, review}.
 Branch III which has sensible thermodynamics and dominates the high temperature physics does not appear to be smooth in this gauge. We also find a curious feature that the spin-3 charge carried by  branch-III vanishes at a special non-zero value of $\mu$, given by $\mu T\,=\,\tfrac{3}{2\pi}$.

A very important aspect of the high temperature study of the system is the identification of the thermodynamical variables appropriate for the ${\cal W}_3^{(2)}$ CFT. Furthermore, it is also essential to confirm the exact form of the deformations of the UV fixed point theory. To this end we perform a Ward identity analysis in the deformed CFT, along the lines of a similar analysis for the IR CFT done in \cite{gutkraus}. It turns out that the equations of motion for the Chern-Simons connections (for non-zero chemical potentials) in a convenient gauge, can be precisely mapped to the Ward identities satisfied by all the currents of the ${\cal W}_3^{(2)}$ CFT deformed by a certain combination of the two spin-$\tfrac{3}{2}$ currents. A crucial element of this map is the identification of the correct energy variable within the Chern-Simons connection, and this is rendered non-trivial by the presence of the spin-1 current in the ${\cal W}_3^{(2)}$ algebra.

The paper is organized as follows: In Section 2, we review the ingredients of the classical black hole solutions in the ${\rm SL}(3,\mathbb R)$ Chern-Simons theory. We also explain how to explicitly compute the correct thermodynamic grand potential and its consistency with the holonomy conditions. In Section 3 we solve the holonomy conditions and exhibit the properties of the different branches with reference to the IR fixed point theory. Section 4 collects some of the salient features of the spacetime/metric description of the different solution branches. Section 5 is devoted to a Ward identity analysis with reference to the UV fixed point, and the emergence of a consistent thermodynamic description of the high temperature black hole solution (branch III). Finally we comment on certain outstanding puzzles and questions for further study in Section 6.

\section{${\rm SL}(3,\mathbb R) \times {\rm SL}(3,\mathbb R)$ Chern-Simons theory} 
We review below the solutions of \cite{gutkraus} where classical black holes in 3d higher spin gravity were first constructed. These were obtained in the ${\rm SL}(3,\mathbb R) \times {\rm SL}(3,\mathbb R)$ Chern-Simons theory corresponding to Einstein gravity coupled to a spin-3 symmmetric tensor field. 

\subsection{Classical solutions}

The classical action for ${\rm SL}(3,\mathbb R) \times {\rm SL}(3,\mathbb R)$ Chern-Simons theory at level $k$ in three dimensions is
\be
I\,=\,I_{\rm CS}[A]-I_{\rm CS}[\bar A]\,,
\ee
with
\be
I_{\rm CS}[A]\,=\,\frac{k}{4\pi}\int{\rm Tr}\left(A\wedge dA +\frac{2}{3} A\wedge A \wedge A\right)\,.
\ee
\\
Following the conventions of \cite{campoleoni, Castro:2011iw} the Chern-Simons level is related to Newton's constant via \footnote{For general ${\rm SL}(N,\mathbb R)$ the corresponding relation is $k=\ell/(8 G\epsilon_N)$, and $\epsilon_N={\rm Tr} L_0^2=\tfrac{1}{12}N(N^2-1)$ \cite{Castro:2011iw}.}
\be
k\,=\,\frac{\ell}{16 G},
\ee
where $G$ is the three dimensional gravitational constant and $\ell$ is the radius of AdS$_3$. The equations of motion for Chern-Simons theory imply that the gauge connections are flat,
\be
F\,=\,dA+A\wedge A\,=0\,,\qquad\qquad\bar F\,=\,d\bar A +\bar A\wedge\bar A\,=0\,.
\ee
From the explicit form of the ${\rm SL}(3,\mathbb R)$ connections, one can deduce the vielbein $e$, and the spin connection $\omega$,
\be
e\,=\,\ell\,\frac{(A-\bar A)}{2}\,,\qquad\qquad \omega\,=\,\frac{(A+\bar A)}{2}\,,
\ee 
and subsequently, the metric and the spin 3 field can be derived as
\cite{campoleoni},
\be
g_{\mu\nu}\,=\,\frac{1}{2}\,{\rm Tr}(e_\mu e_\nu)\,,\qquad\qquad
\varphi_{\mu\nu\lambda}\,=\,\frac{1}{3!} \,{\rm Tr}(e_\mu e_\nu e_\lambda)\,.
\ee
We work in Euclidean signature with boundary coordinates $z=i\,t +\phi$  and $\bar z = i\,t -\phi$. The angular variable $\phi$ has period $2\pi$ whilst the periodicity of Euclidean time is set by the inverse temperature, $\beta$. Consequently, the holomorphic coordinate $z$ is identified under shifts
$z\simeq z+2\pi\tau\,$, where $\tau\equiv i\beta/2\pi$.
 
The flat connections corresponding to the {\em non-rotating} ${\rm SL}(3,\mathbb R)$ black hole in ${\rm AdS}_3$, carrying a spin-3 charge, can be written in terms of gauge equivalent constant connections $a$ and $\bar a$, which are in turn functions of the generators $L_0, L_{\pm 1}, W_0,W_{\pm 2}\,$:
\bea
A\,=\,b\,a\, b^{-1} + bdb^{-1}\,,\qquad
\bar A\,=\,b^{-1}\bar a\, b + b^{-1}db\,,\qquad b=e^{- \rho L_0}
\eea
and
\bea
&&a\,=\, \left(L_1 -\frac{\pi}{2k}\,{\cal L}\,L_{-1}-\frac{\pi}{8k}\,{\cal W}\,W_{-2}\right)\,dz\label{connections}\\\nonumber
&&\hspace{1.8in}+\,\mu\,\left(W_2-\frac{\pi}{k}\,{\cal L}\,W_0+\frac{\pi^2}{4k^2}\,{\cal L}^2\,W_{-2}+\frac{\pi}{k}\,{\cal W}\,L_{-1}\right)\,d\bar z\,,\\\nonumber\\\nonumber
&&\bar a\,=\,\mu\,\left(W_{-2}-\frac{\pi}{k}\,{\cal L}\,W_0+\frac{\pi^2}{4k^2}\,{\cal L}^2\,W_{+2}+\frac{\pi}{k}\,{\cal W}\,L_{+1}\right)\,dz\\\nonumber
&&\hspace{3in}- \left(L_{-1} -\frac{\pi}{2k}\,{\cal L}\,L_{1} -
\frac{\pi}{8k}\,{\cal W}\,W_{2}\right)\,d\bar z\,.
\eea
Here, ${\cal L}$ and ${\cal W}$ are the ``charges'' conjugate to the thermodynamic potentials $\beta$ and $\mu$. Up to a normalization factor they are, respectively, the mass and  spin-3 charge of the solution. The form of the connections above differs slightly from that of \cite{gutkraus}, 
and we recover the latter upon making the replacement $k\to k/4$. As discussed previously, this is because we have chosen to set $k\,=\,{\ell}/{16 G}$.

\subsection{Holonomy conditions and free energy}
\label{thermoaction}

In ordinary gravity, the time circle shrinks smoothly at the horizon of a Euclidean black hole. For the 3d higher spin theory, the analogous condition is encoded in the holonomy of the ${\rm SL}(3,\mathbb R)$ gauge field around the Euclidean time direction \cite{gutkraus}. Specifically, a smooth solution requires this holonomy to be trivial:
\\
\bea
{\rm Hol}_t(A)\,\equiv\,{\cal P}\exp\left(\oint A_t\right)\,=\,{\bf 1}\,,\qquad{\rm Hol}_t(\bar A) = {\bf 1}\,.
\eea
Equivalently, this requires that the eigenvalues of the matrix $\,\beta \,a_t\,$ be given by $(2\pi i,\, 0,\, -2\pi i)\,$, so that
\be
{\rm det}(\beta\,a_t)\,=\,0\,,\qquad{\rm and}\qquad{\rm Tr}(\beta^2\, a_t^2)\,=\,-\,8\pi^2\,.
\ee
Expressed in terms of the potentials,
\be
\tau \equiv \frac{i\,\beta}{2\pi}\,,\qquad\qquad
\alpha \equiv -\frac{i\,\beta}{2\pi}\,\mu\,,
\ee
the two holonomy conditions yield (non-linear) algebraic relations between the pair $(\tau,\alpha)$ and the conjugate charges $({{\cal L}, {\cal W}})$:
\bea
&&{\cal W}\,=\,-\frac{\tilde k}{24 \pi \alpha\,\tau}-\frac{\tau\,{\cal L}}{3\,\alpha}
-\frac{32\pi{\cal L}^2\,\alpha}{ 9\tau\tilde k}\,,\qquad\qquad \boxed{\, \tilde k \equiv4\,k\,\,}\label{conditions}\\\nonumber\\\nonumber
&&27\,{\tilde k}^{2}\,{\cal W}\,\tau^3+576\,\tilde k \pi\,{\cal L}^2 \alpha\tau^2+
864\pi \tilde k\, {\cal L}{\cal W}\,\alpha^2\tau -2048 \pi^2 {\cal L}^3\alpha^3 +864\pi\tilde k\,{\cal W}^2\alpha^3\,=\,0\,.
\eea
Here we have chosen to write the result in terms of $\tilde k = 4k$, to make the point that the conditions are identical to the expressions of \cite{gutkraus}, up to a rescaling of the Chern-Simons level.
It is then a straightforward excercise to check that the energy, ${\cal L}$ and the higher spin charge ${\cal W}$, regarded as functions of $\tau$ and the chemical potential $\alpha$, satisfy the integrability condition,
\be
\frac{\partial {\cal L}}{\partial\alpha}\,=\,\frac{\partial {\cal W}}{\partial \tau}\,.
\ee
\paragraph{\underline{On-shell action}:}
This implies the existence of a thermodynamical action such that
\bea
dI_{\rm th} \propto \left\{{\cal L}(\tau,\alpha)\,d\tau +{\cal W}(\tau,\alpha)\,d\alpha\right\}\,.\label{thermod}
\eea
In principle, this thermodynamical action can be deduced by integrating the smoothness/holonomy conditions. At first sight, however, this appears cumbersome. Instead, we turn to a direct evaluation of the thermodynamical action following the proposal of \cite{bct} using the form of the connections alone. According to this proposal, the full {\em on-shell} action can be evaluated by using ``angular quantization'', i.e. foliating the bulk solid torus by disks at constant $\phi$. The bulk contribution to the Chern-Simons action vanishes on-shell, leaving behind only a boundary term at infinity,
\be
 I_{\rm on-shell}\,=\,-\frac{k}{4\pi}\int_{{\mathbb T}^2} dt\,d\phi\,{\rm Tr}(a_t\,a_\phi)\quad
-\quad(a\to \bar a)\,.
\label{boundary}
\ee
Evaluating it for the ${\rm SL}(3,\mathbb R)$ connections \eqref{connections} we find
\be
 I_{\rm on-shell}\,=\, -2\pi{\cal L}\,\beta\,+\,\frac{16\pi^2\,\beta {\cal L}^2 \mu^2}{3\,k}\,.
\ee
\paragraph{\underline{Thermodynamical action:}}
Pinning down the correct thermodynamical action requires some more work. We will also see below that we need to introduce a small shift in both the
on-shell action and the thermodynamical action following from the proposal of \cite{bct}, in order to be consistent with the holonomy conditions \eqref{conditions}. For the moment, we  follow the analysis of \cite{bct} and first determine the 
variation of the Chern-Simons action, treated as a function of ${\beta,\mu,{\cal L}}$ and ${\cal W}$, which reduces to a boundary term
\be
\delta I (\beta,\mu,{\cal L},{\cal W})\,=\,\frac{k}{4\pi}\int_{{\mathbb T}^2}dt\,d\phi\,{\rm Tr}(a_\phi\,\delta a_t-a_t\,\delta a_\phi)\quad-\quad(a\to\bar a)\,.
\ee
The variations $\delta a_t, \delta a_\phi$ are easy to compute using \eqref{connections} and, importantly, are deduced by allowing all thermodynamic parameters to vary freely. We then find,
\be
\delta I\,=\,-2\pi d{\cal W}\,\beta \mu\, - 6 \pi {\cal W}\,\beta\,d\mu +d\beta\left(2\pi {\cal L}\,-\,\frac{16\pi^2\,{\cal L}^2 \mu^2}{3\,k}\right)
\ee
It is now clear that to obtain a  thermodynamical action which is only a function of the 
potentials $(\beta,\mu)$ or $(\tau,\alpha)$ alone, we can simply perform a Legendre transform of $\tilde I$:
\be
\tilde I_{\rm th}\,=\, I_{\rm on-shell}\, +\,2 \pi{\cal W}\,\beta\mu\,,
\ee 
so that
\be
\delta \tilde I_{\rm th}\,=\,\left(2\pi{\cal L}\,-\,\frac{16\,\pi^2\,{\cal L}^2 \mu^2}{3\,k} + 2\pi{\cal W}\,\mu \right)\, d\beta - 4\pi{\cal W}\,\beta d{\mu}\,.
\ee
Note that we use the symbol $\tilde I_{\rm th}$ for the {\em proposed} thermodynamical action. We will see below that it needs to be modified slightly to make it compatible with eq.\eqref{thermod}. To check this condition, we rewrite $\tilde I_{\rm th}$ and its variation, in terms of the potentials $(\tau,\alpha)$:
\be
\tilde I_{\rm th}\,=\,-2\pi i\,\left(-2\pi {\cal W}\,\alpha - 2\pi{\cal L}\,\tau
+\frac{16\pi^2{\cal L}^2\alpha^2}{3 k\, \tau}\right)
\ee
Making use of the holonomy conditions \eqref{conditions} 
to compute the derivatives $\partial{\cal L}/\partial{\alpha}$ and $\partial {\cal L}/\partial\tau$, we find
\be
\frac{i}{2\pi}\,\frac{\partial{\tilde I}_{\rm th}}{\partial\tau}\,=\, 4\pi{\cal L} +\frac{k}{\tau^2}\,,\qquad
\frac{i}{2\pi}\,\frac{\partial{\tilde I}_{\rm th}}{\partial\alpha}\,=\, 4\pi\,{\cal W}\,.
\ee
(Since we are only interested in non-rotating solutions, the energies of the two chiral sectors are equal, ${\cal L} =\bar {\cal L}$ and ${\cal W} = -\bar{\cal W}$). 
Therefore the correct thermodynamic action satisfying the usual thermodynamic relations, and which is consistent with the holonomy conditions is,
\be
\boxed{I_{\rm th}\,=\,\tilde{I}_{\rm th} - \frac{2\pi i\,k}{\tau}
}\,. \label{shift}
\ee
Now we can write down the thermodynamic grand potential
$\Phi\equiv T I_{\rm th}$ where $T\equiv \beta^{-1}$ \, and the holonomy equation \eqref{conditions} has been used to cast the result in an especially simple form, 
\be
\boxed{\,\,\Phi\,=\, -\, 4\pi{\cal L}\, +\, 8 \pi{\cal W}\,\mu \,\,\,}.
\ee
The apparent simplicity of this expression hides the complicated dependence of the charges ${\cal L}$ and ${\cal W}$ on the temperature $T$, and chemical potential $\mu$.

It is worth noting that the shift \eqref{shift} that we had to introduce to obtain consistent thermodynamics, would also be necessary in the limit $\mu\to0$ to yield the correctly normalized free energy of the BTZ black hole. It is not clear to us how to understand or interpret its origin in an independent fashion. The entropy  follows directly from the grand potential via $S = (4\pi {\cal L}- 4\pi{\cal W}\,\mu-\Phi)/T$,
\be
\boxed{\,\,S\,=\,\frac{1}{T}\,\left(\,8\pi{\cal L}\,-\,12\pi{\cal W}\,\mu\,\right)\,\,}\,,\label{entropy}
\ee
and satisfies the check that $S = -\partial\Phi/\partial T$.
To make contact with the expressions of \cite{gutkraus,review} we first solve the first holonomy condition to obtain $T= \sqrt{\tfrac{\cal L}{2\pi k}}\,(1- 3\mu\tfrac{\cal W}{\cal L} + \tfrac{8}{3k}\,\pi\mu^2{\cal L})^{1/2}$ and subsequently use the second holonomy condition to solve for $\mu$, to yield
\be
S\,=\, 8\pi\sqrt{2\pi{\cal L}\,k}\,f(y)\,,\qquad\qquad y\,=\,\frac{27 \,k\,{\cal W}^2}{16\pi\,{\cal L}^3}.\label{yentropy}
\ee 
The appearance of the dimensionless variable $y\sim {\cal W}^2/{\cal L}^3$ is natural and dictated by dimensional analysis.

\section{Multiple branches}

The original work of \cite{gutkraus} and subsequent papers on the topic 
\cite{ammon, Kraus:2011ds} were focussed on solving the holonomy conditions \eqref{conditions} subject to the requirement that the higher spin charge ${\cal W}$ vanishes with the chemical potential $\mu$. However, the equations \eqref{conditions} admit three additional solutions. These new branches have infinite free energy in the $\mu\to 0$ limit, but for any finite $\mu$ they can compete in the thermal ensemble with the black hole solution discussed in  \cite{gutkraus}.

To understand the features of the solutions of Eq.\eqref{conditions}, we begin by noticing that the level $k$ can be eliminated from the equations by performing the rescalings
\be
\mu \to \mu\sqrt{\tilde k}\,,\qquad T\to \frac{T}{\sqrt {\tilde k}}\,,\qquad {\cal W}\to \frac{\cal W}{\sqrt {\tilde k}}\,,\qquad {\cal L} \to {\cal L}\,,
\ee
where $\tilde k = 4k$.
We can reintroduce $k$ into any subsequent expression by reversing the process.
We first substitute the expression for ${\cal W}$ into the second holonomy condition to obtain a quartic equation for the energy ${\cal L}$,
\bea
&&65536\pi^3\,\mu^6\,{\cal L}^4-18432\pi^2\,\mu^4\,{\cal L}^3 + (1728\pi\mu^2-6144\pi^3 T^2 \mu^4)\,{\cal L}^2\,+ \\\nonumber
&&+\,(288\pi^2\,T^2\,\mu^2-54)\,{\cal L}\,+\,27\pi\,T^2\,+\,144\pi^3\,T^4\,\mu^2\,=\,0\,.
\eea

\paragraph{\underline {Low $T$ solutions:}} By dimensional analysis of the holonomy condition, we learn that the combination 
${\cal L}/T^2$ is only a function of the dimensionless variable $\mu T$. The parameter $\mu$ is a  coupling constant with dimension $-1$ as it acts as a chemical potential for the spin-3 charge whose corresponding current is a dimension-3 operator.
For present purposes we define `low' and `high' temperatures as $\mu T\ll 1$ and $\mu T \gg 1$ respectively. Since the Chern-Simons theory is being treated classically, we always have $k\gg 1$, and to be able to describe classical black hole like solutions we must also have the (weak) requirement $T\gg k^{-1}$.

Examining the quartic for small $\mu\,T$, we infer the existence of four branches of solutions. 
`Branch I', according to our nomenclature, reduces to the BTZ solution when $\mu, {\cal W}\to 0$, and we will also refer to this as the `BTZ-branch'.  Branches II and III are  new solutions which are thermodynamically unstable and stable, respectively, whilst Branch IV seems to be an unphysical solution with negative entropy. 

At $T=0$, the equation simplifies to ${\cal L}\,(3-32\pi\,\mu^2 {\cal L})^3=0$, so that there is one zero energy solution (the BTZ-branch) and three other coincident roots with finite mass. For small finite $T$, the coincident roots split. The low temperature properties of the four solutions are summarized in Table \eqref{table:lowt}.
\begin{table}[ht] 
\centering 
\begin{tabular}{|c| c| c| c|} 
\hline
Branch & $4\pi{\cal L}/k$ & $S/k$ & $\Phi$ \\ 
[1ex]	
\hline 
&&&
\\
I& $8\pi^2\,T^2 +\tfrac{640}{3}\pi^4 \mu^2 T^4$ & $ 16\pi^2 T\,+\tfrac{512}{3}\pi^4 \mu^2 T^3 $&
$-8\pi^2 T^2\,k -\tfrac{128}{3}\pi^4\mu^2 T^4k$
\\ & & &  
\\ II& $ \tfrac{3}{2}\,\mu^{-2} - 6\pi\tfrac{T}{\mu}\ldots$ &
$ 6 \pi \mu^{-1}-\tfrac{32}{3}\pi^2 T  $ & $\tfrac{k}{2}\mu^{-2} -6 k\pi\tfrac{T}{\mu}\ldots$
\\ & & &
\\ III & $ \tfrac{3}{2}\mu^{-2} + \tfrac{8}{3}\pi^2\,T^2\ldots$ & $\tfrac{16}{3}\pi^2 T-\tfrac{512}{243}\pi^4\mu^2T^3\ldots $ & $\tfrac{k}{2}\mu^{-2} -\tfrac{8}{3}k\pi^2 T^2\ldots$
\\ & & &
\\ IV & $ \tfrac{3}{2}\mu^{-2} + 6\pi\tfrac{T}{\mu}\ldots$ &$-6 \pi\mu^{-1}-\tfrac{32}{3}\pi^2 T\ldots$  
& $\tfrac{k}{2}\mu^{-2} +6k\pi\tfrac{T}{\mu}\ldots$
\\ 
 [1ex] 
\hline 
\end{tabular} 
\caption{\small Low $T$ energy, entropy and free energy of the four solutions. The Chern-Simons level 
and the central charge of the ${\cal W}_3$ CFT are related as 
$k\,=\,c/24$.}
\label{table:lowt} 
\end{table}

It is easily checked that all solutions carry non-zero spin-3 charge ${\cal W}$. In particular, the BTZ-branch has vanishing ${\cal W}\approx \tfrac{64}{3}\pi^3\mu\, T^4\,k$ at zero temperature, whilst all the other solutions have ${\cal W} \approx k/(4\pi\mu^3)$. Since the higher spin charge of the new solutions diverges at small $\mu$, one may be tempted to discard them. However, for any non-vanishing $\mu$, three of the four branches appear to be solutions with physically acceptable properties, and we should therefore look for an interpretation of these new branches.

Although branch III is thermodynamically disfavoured, like the BTZ-branch it has an entropy that increases linearly with $T$. Furthermore, the entropy of these solutions vanishes at zero temperature even though the higher spin charge remains non-zero.

\paragraph{\underline {High $T$ solutions:}} Interestingly, for large temperatures and/or chemical potential $\mu\,T\gg1$, the holonomy conditions admit only two real roots. These correspond to branches III and IV, the latter with negative entropy. In particular, the BTZ-branch has disappeared. The asymptotic behaviour of the energy and entropy of the two real branches at high temperature are:
\bea
&&{\tfrac{4\pi}{k}{\cal L}}_{\rm III}\to 2\sqrt 3\,\pi\,\tfrac{T}{\mu}\,-\,3^{3/4}\sqrt{\tfrac{\pi}{2}}\,\tfrac{\sqrt T}{\mu^{3/2}} ,\qquad\quad  
\tfrac{1}{{k}}S_{\rm III}\to 4\sqrt2\, 3^{1/4}\pi^{3/2}\,\sqrt{\tfrac{ T}{\mu}}- \tfrac{2\sqrt 3 \,\pi}{\mu}\,,
\nonumber\\\label{highT}\\\nonumber
&&{\tfrac{4\pi}{k} {\cal L}}_{\rm IV}\,\to 2\sqrt 3\,\pi\tfrac{T}{\mu}+3^{3/4}\sqrt{\tfrac{\pi}{2}}\,\tfrac{\sqrt T}{\mu^{3/2}}\,,\qquad\quad 
\tfrac{1}{{k}}{S_{\rm IV}}\,\to -\,4\sqrt2\, 3^{1/4}\pi^{3/2}\sqrt{\tfrac{ T}{\mu}} - \tfrac{2\sqrt 3 \,\pi}{\mu}\,.
\eea
Whilst both these solutions have positive specific heats, the temperature dependence of their energies and entropies poses a puzzle. The negative entropy of branch IV suggests that the solution may be unphysical. However, the entropy of branch III, while positive, has an unusual dependence on temperature. Instead of scaling linearly with temperature as expected for a (dual) 2d CFT, it scales as $\sim \sqrt T$. Similar comments apply to the temperature dependence of the energy ${\cal L}$ as measured with respect to the stress tensor of the ${\cal W}_3$ CFT.

\begin{figure}[h]
\epsfig{file=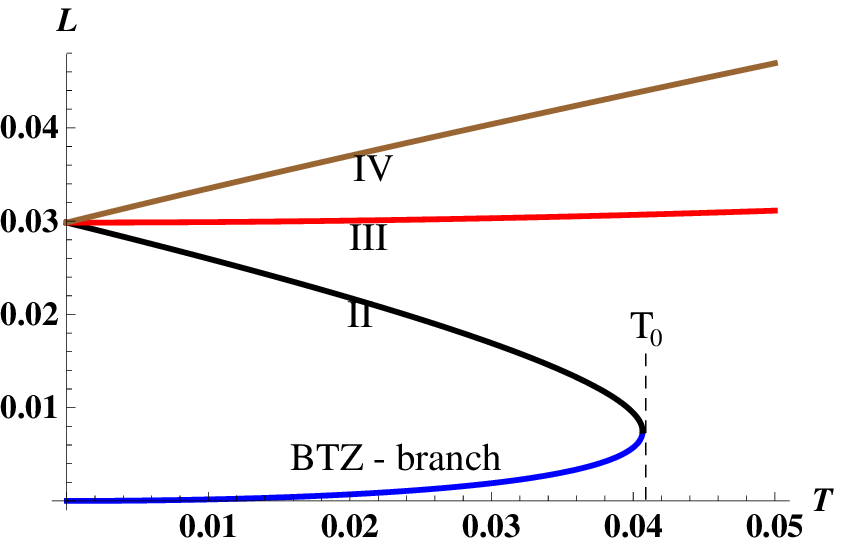, width =2.4in}\hspace{1in}
\epsfig{file=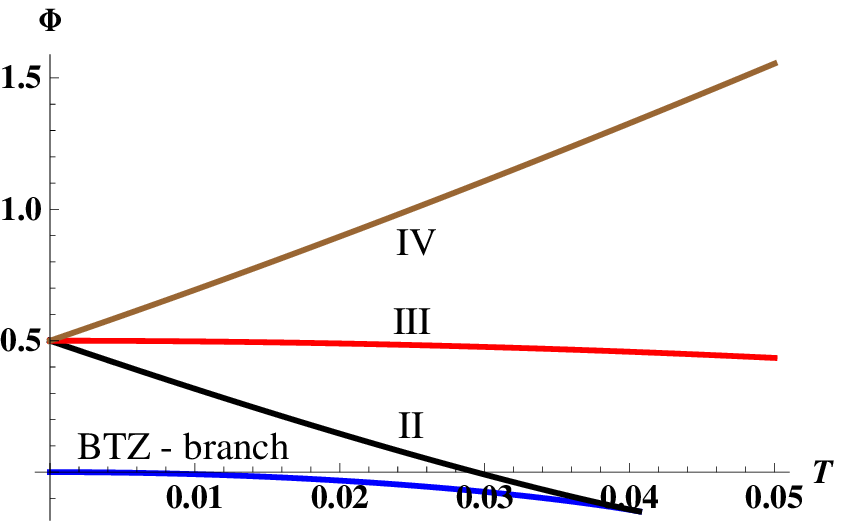, width =2.4in}\\\\\\
\vspace{0.4in}
\epsfig{file=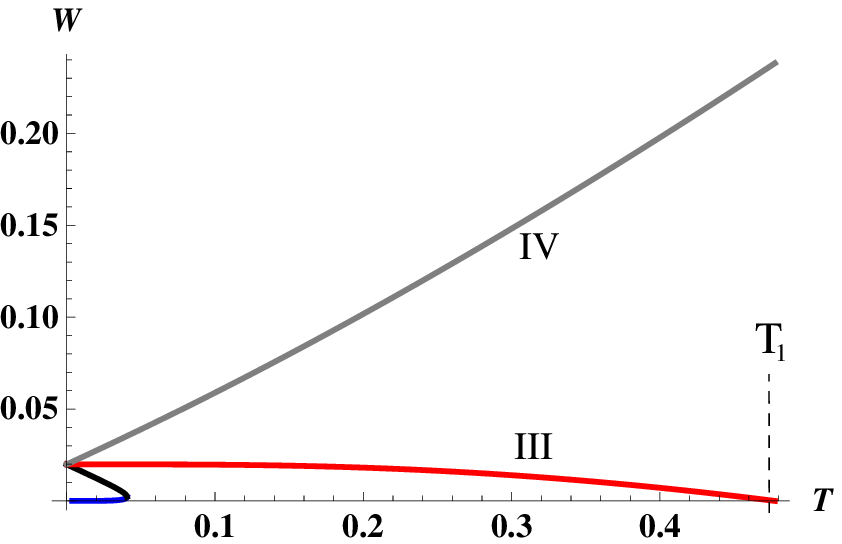, width =2.4in}\hspace{1in}
\epsfig{file=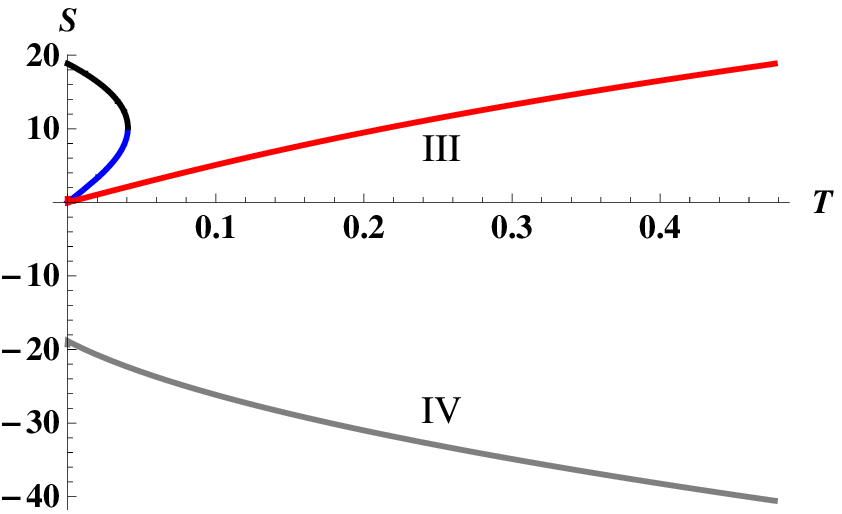, width =2.4in}
\caption{\small{ Thermodynamics of the four solutions to the holonomy/smoothness conditions, each carrying non-zero spin-three charge, plotted numerically for $\mu=1$.}}
\label{fig2} 
 \end{figure}

The complete situation is depicted in Figure\eqref{fig2}.
The plots display two striking features:
\begin{itemize}

\item{The first is the merger of the BTZ-branch with the thermodynamically unstable branch II which occurs at a specific temperature (for a given $\mu$). In fact we can pin-point the temperature $T_0$, at which this occurs by computing the discriminant of the quartic polynomial (in ${\cal L}$) and setting it to zero. We find,
\be
\mu\,T_0\,=\,\frac{3}{16\pi}\,\sqrt{2\sqrt 3 - 3}\,.
\ee
The mass and spin-3 charge of the black holes at this point are 
\be
{\cal L}\,=\,\tfrac{8}{9}\pi\,T_0^2\,(3+2\sqrt 3)\,k\,,
\qquad {\cal W}\,=\,\frac{64\,(3+5\sqrt 3)}{81\sqrt{2\sqrt 3 -3 }}\,\pi^2\,T_0^3\,k\,.
\ee
The main point to note is that for temperatures $T>T_0$, the roots associated to these two branches move into the complex plane and the number of real solutions is reduced to two. Of the two remaining solutions one appears to be unphysical as it has negative entropy and higher free energy, so that branch III appears to be the only thermodynamically stable, classical configuration dominating the thermal ensemble.
Taken seriously, this picture would imply a discontinuous transition at 
$T = T_0$, where the free energy of the system jumps and the 
specific heat diverges.}

\item{The second striking feature is that at a higher temperature $T_1 > T_0$, the stable high temperature solution (branch III) has vanishing spin-3 charge (with $\mu\neq 0$). Once again it is possible to determine this temperature quite easily and we find,
\be
\mu\,T_1\,=\,\frac{3}{2\pi}\,,\qquad\qquad {\cal L}_{\rm III}\,=\,\tfrac{1}{2}\,\pi T_1^2\,k\,,\qquad {\cal W}\big|_{T_1}\,=\,0\,.
\ee
The existence of this point is particularly counter-intuitive 
since it implies a new uncharged black hole solution in the higher spin theory (albeit deformed by the chemical potential $\mu$), in addition to the BTZ back hole.}
\end{itemize}
For higher temperatures $T>T_1$, the charge ${\cal W}$ of this solution (branch III) turns negative, and it remains the only classical saddle point dominating the ensemble. The rest of the paper will be devoted to piecing together properties of this branch of solutions and in particular, establishing their physical significance, if any. 

\section{Spacetime geometry}

The introduction of the chemical potential for spin-3 charge corresponds to a deformation of the ${\cal W}_3$ CFT by an irrelevant operator. This alters the asymptotics of the dual background (in a fixed gauge), so that the UV geometry is actually ${\rm AdS}_3$ spacetime with radius $\ell/2$:
\be
\frac{1}{\ell^2}\,ds^2\big|_{\rho\to\infty}\,=\,  (4\mu^2\,e^{4\rho}+e^{2\rho}+\ldots)\,dt^2\,+\,(4\mu^2\,e^{4\rho}+e^{2\rho}+\ldots)\,d\phi^2\,+\,d\rho^2\,.
\ee
The growth of $g_{tt}$ and $g_{\phi\phi}$ as $e^{4\rho}$ instead of $e^{2\rho}$, signals an asymptotic ${\rm AdS}_3$ with radius $\ell/2$. The altered asymptotics is due to the appearance of the generators $W_{\pm 2}$ of 
${\rm SL}(3,\mathbb R)$ in the connections \eqref{connections}. The ${\rm SL}(2,\mathbb R) \times {\rm SL}(2,\mathbb R)$ isometry of this new AdS geometry is generated by two copies of the set $\{\pm\tfrac{1}{4}W_{\pm 2},\,\tfrac{1}{2} L_0 \}$, constituting a different embedding of $sl(2,\mathbb R)$ in $sl(3, \mathbb R)$. As we will briefly review in Section 5, this leads to the Polyakov-Bershadsky ${\cal W}_3^{(2)}$ asymptotic isometry.

An alternate way to view this flow is as a relevant deformation of the ${\cal W}_3^{(2)}$ CFT by an operator with scaling dimension $3/2$. To see this it is more natural to rewrite the asymptotic metric (after shifting $\rho$) as
\bea
&&\frac{1}{\ell^2}\,ds^2\,=\, + (e^{4\rho}+ \lambda^2\,
e^{2\rho}+\ldots)\,dt^2\,+\,(e^{4\rho} + \lambda^2\, e^{2\rho}+\ldots)\,d\phi^2\,+\,d\rho^2\,,\\\nonumber
&&\lambda\,\equiv\,\frac{1}{2\sqrt\mu}\,.
\eea
At the UV scale invariant point, $\lambda$ has  scaling dimension 
$\tfrac{1}{2}$, and is the source for a dimension $\tfrac{3}{2}$ operator. This is also completely consistent with the relation between $\lambda$ and $\mu$, the latter being a constant with scaling dimension $-1$. Indeed the RG flow background in the absence of spin-3 charge and energy density is an exact solution of the ${\rm SL}(3,\mathbb R)$ Chern-Simons theory,
\bea
&&ds^2\,=\,d\rho^2 - \left(\tfrac{1}{4}\,e^{4\rho} +\lambda^2\, e^{2\rho}\right)\,dz\,d\bar z\,,\\\nonumber\\\nonumber
&& \varphi_{\alpha\beta\gamma}\,dx^\alpha\,dx^\beta\,dx^\gamma=\,\lambda^2\,e^{4\rho}\,\left(d\bar z^3-dz^3\right)\,.
\eea
Although this solution does not carry higher spin charge, spin-3 fields are turned on in the background.
With non-vanishing spin-3 charge and energy density, the spacetime metric generically acquires the form of a wormhole geometry wherein both spatial and temporal circles remain finite everywhere.
It was, however, shown in \cite{ammon} that one can find an ${\rm SL}(3,\mathbb R)$ transformation which turns the wormhole gauge metrics to black hole spacetimes with smooth horizons. We refer the reader to \cite{ammon, review} for the details of the construction of these gauge transformations. We will only quote their results and focus on the nature of the black hole geometries associated to some of the new branches discussed above.

\subsection{Branch II and the extremal black hole}
From our thermodynamical analysis we have seen that there is precisely one branch (namely branch II) of classical solutions to the holonomy conditions, which has non-zero entropy and spin-3 charge as the zero temperature, ``extremal'' limit is approached. In \cite{ammon,review} a black hole gauge geometry with exactly this property was obtained. Indeed, rewriting the entropy of branch II near $T=0$ in microcanonical variables yields 
\be
S_{\rm II}\,=\,k\,\left(\frac{6\pi}{\mu} - \frac{32}{3}\pi^2\,T +\ldots\right)\,=\, 8\pi\sqrt{2\pi k{\cal L}}\,\left(\frac{\sqrt 3}{2} + \frac{\sqrt{2-y}}{6\sqrt 2}+\ldots\right)\,,
\ee
which correctly matches the entropy of the extremal black hole of \cite{ammon,review}. We have seen already that the specific heat of  
this branch is negative and particularly at $T=0$,
\be
\lim_{T\to 0}\,4\pi\,\frac{\partial { {\cal L}_{\rm II}}}{\partial T}\,=\,
-6\pi k\,\mu^{-1}\,,
\ee
which renders the extremal solution thermodynamically unstable. 

It is useful to see the metric correspondng to the connections 
\eqref{connections} in black hole gauge deduced in \cite{ammon},
\bea
&& g_{rr}\,=\,\frac{(C-2)(C-3)}{(C-2-\cosh^2 r)^2}\label{bhgauge}\\\nonumber
&& g_{tt}\,=\,- \left(\tfrac{8\pi\,{\cal L}}{\tilde k}\right)\left(
\frac{C-3}{C^2}\right)
\frac{(a_t + b_t\,\cosh^2 r)}{(C-2-\cosh^2 r)^2}\,\sinh^2 r\\\nonumber
&& g_{\phi\phi}\,=\,\left(\tfrac{8\pi\,{\cal L}}{\tilde k}\right)\left(
\frac{C-3}{C^2}\right)\frac{(a_\phi + b_\phi\,\cosh^2 r)}{(C-2-\cosh^2 r)^2}\,\sinh^2 r + \left(\tfrac{8\pi\,{\cal L}}{\tilde k}\right) (1+\tfrac{16} {3}\gamma^2 + 12 \gamma\zeta)\,,
\eea
where the parameters $\zeta$, $C$ and $\gamma$ are defined as
\be
\zeta\,=\sqrt{\frac{\tilde k\, {\cal W}^2}{32\pi {\cal L}^3}}\,,\qquad
\gamma\,=\,\sqrt{\frac{2\pi {\cal L}}{\tilde k}}\,,
\qquad\zeta\,=\,\frac{C-1}{C^{3/2}}\,,\qquad \tilde k = 4k\,
\ee
The parameters $a_t, b_t, a_\phi$ and $b_\phi$ are functions of $\gamma$ 
and $C$, and are listed in  Appendix\eqref{appb}. 
The horizon of the geometry is at $r=0$ and the UV asymptotics (${\rm AdS}_3$ with radius $1/2$) emerges when the denominators are vanishing, near $r=r^*$ where $\cosh^2 r^*\,=\,C-2$. It was shown in \cite{ammon} that requiring smoothness of the Euclidean metric at the horizon yields precisely the holonomy conditions eqs.\eqref{conditions}.

The extremal limit that we have discussed occurs when $\zeta\to \frac{2}{3\sqrt 3}$ or equivalently $C\to 3$. In this limit, the metric seemingly degenerates while the asymptotic region, now near $r^*=0$, also approaches the horizon at $r=0$. The limiting metric can actually be obtained by perfoming a coordinate rescaling that effectively stretches the region between these two. We will not repeat that analysis here.

The main point to be made here is that the black hole gauge metric remains smooth and describes both branches I and II shown in Figure\eqref{fig2}. The two branches are covered by the range 
\be
3\leq C \leq \infty\,,
\ee
the upper limit corresponding to the (uncharged) BTZ black hole. The merger of the two branches occurs when $C= 3 + \tfrac{3}{2}\sqrt 3$, which falls within the range above. It appears that both the wormhole gauge and black hole gauge metrics remain smooth and unremarkable at this merger point.

\subsection{Branches III and IV and spacetime metric}
The black hole gauge metric above does not seem to be a useful or suitable description of branches III and IV.  Both these branches appear to occupy the parametric range $C<3$, where either the signature of the metric 
\eqref{bhgauge} is unphysical with $g_{rr} < 0$ or the asymptotic ${\rm AdS}_3$ is not accessible (when $C<2$). There should exist a different gauge transformation that can turn the wormhole gauge metric for these branches into a smooth metric with a horizon and sensible asymptotics. 
Regardless, the wormhole gauge  spacetime metric is well defined. Since branch IV appears to be disfavoured thermodynamically for all temperatures, and its entropy (as measured with reference to the IR ${\cal W}_3$ CFT) is negative, we will not discuss it further.

Branch III which appears to be physical, has the curious property
that its spin-3 charge vanishes at $\mu\,T_1\,=\,\tfrac{3}{2\pi}$. Remarkably, using the connections from \eqref{connections} (with ${\cal W}=0$), we find that the metric in wormhole gauge acquires a ``horizon'' i.e. as $T\to T_1$, the neck of the wormhole pinches off:
\bea
&&g_{\rho\rho}\,=1\,,\\\nonumber\\\nonumber
&&g_{tt}\big|_{T=T_1}\,=\, -\,\frac{1}{16\pi^2\,T^2}\,
\left(e^{2\rho}\,-\,\frac{\pi^2 T^2}{4}  \right)^2\,\left(144\,\,+\,88\pi^2\,T^2\,e^{-2\rho}+\,9\pi^4\,T^4\,e^{-4\rho}\right)\,,\\\nonumber\\\nonumber
&&g_{\phi\phi}\big|_{T=T_1}\,=\, \frac{9}{\pi^2\,T^2}\,e^{4\rho}\,+\,e^{2\rho}\,+\,\frac{19\pi^2 T^2}{8}\,+\,\frac{\pi^4 T^4}{16}\,e^{-2\rho}\,+\,\frac{9\pi^6 T^6}{256}\,e^{-4\rho}\,.
\eea
Therefore $g_{tt}$ has a double zero, or a horizon at
\be
\rho \,=\,\rho_h\,=\,\ln\left(\frac{\pi T_1}{2}\right)\,.
\ee
In Euclidean signature, we find that the horizon is actually non-smooth and that there is a conical excess. We have no reason to believe that this is potentially problematic: in particular, it is possible that one could find a gauge transformation that removes this apparent singularity. Alternatively, if this were truly a singular solution we would expect there to be a gauge-invariant characterization of the singular behaviour. However, all physical observables of the Chern-Simons theory, such as the energy, entropy and free energy vary smoothly across this point.

\section{The ${\cal W}_3^{(2)}$ CFT and branch III}

The analysis of the thermodynamics of branch III posed certain puzzles. Specifically, the dependence of the energy and the entropy on temperature, for $\mu T\gg 1$, is not what one would expect from a 2d CFT at high temperature.
On the other hand, we know that the Chern-Simons connection \eqref{connections} represents a flow from the ${\cal W}_3$ CFT in the IR to the so-called ${\cal W}_3^{(2)}$ CFT in the UV. Therefore in the high temperature limit we should expect to be probing properties of this UV CFT at finite temperature. In particular, the branch III of higher spin black hole solutions should exhibit thermodynamics compatible with that of the UV conformal field theory. We now attempt to address this puzzle.

\subsection{The ${\cal W}_3^{(2)}$ algebra:}
There are two inequivalent ways of embedding the ${\rm SL}(2,\mathbb R)$ algebra in 
${\rm SL}(3,\mathbb R)$. The principal embedding can be shown to give rise to the asymptotic ${\cal W}_3$ algebra using the classical Drinfeld-Sokolov procedure \cite{campoleoni, Henneaux:2010xg}. The other `diagonal' embedding yields the so-called ${\cal W}_3^{(2)}$ algebra which is generated by the stress tensor $T(z)$, two spin-$\tfrac{3}{2}$ currents $G^\pm(z)$ and one spin-1 current $J(z)$. The OPE's and the algebra, also referred to as the Polyakov-Bershadsky algebra \cite{Polyakov:1989dm, Bershadsky:1990bg} are listed in Appendix \eqref{appa} \footnote{See, for example, \cite{wyllard} for more details.}.
By directly comparing the ${\cal W}_3^{(2)}$ algebra \eqref{w32}, and the global part of the ${\cal W}_3$ algebra, we infer that the global part of the former is generated by $\hat L_0, \hat L_1, \hat L_{-1}, G^{\pm}_{\pm 1/2}, J_0$, and that these are related to the ${\rm SL}(3,\mathbb R)$ generators as,
\bea
&&\hat L_0\,=\,\tfrac{1}{2}L_0\,,\qquad\qquad
\hat L_{\pm 1}\,=\,\pm\tfrac{1}{4}\,W_{\pm 2}\,,\qquad\qquad 
J_0\,=\,\tfrac{1}{2}W_0\,,\label{redef}\\\nonumber\\\nonumber
&&G^{\pm}_{1/2}\,=\,
\tfrac{1}{\sqrt 8}\,(W_1\mp L_1)\,,\qquad\qquad
G^{\pm}_{- 1/2}\,=\,
\tfrac{1}{\sqrt 8}\,(L_{-1}\pm W_{-1})\,.
\eea
This relation to the ${\rm SL}(3,\mathbb R)$ algebra requires half-integer moding (Neveu-Schwarz b.c.) for the spin-$\tfrac{3}{2}$ currents. If one considers the Ramond sector (integer moding for $G^\pm$), then the global part of the algebra (without a central term) is not $sl(3, \mathbb R)$. For this reason we do not discuss the integer moded case. 

\subsection{Ward identities for deformed ${\cal W}_3^{(2)}$ CFT}

Let us now consider the ${\rm SL}(3,\mathbb R)$ connections $(a,\bar a)$ for the higher spin black hole,  written in a form suitable for discussing the asymptotic $W_{3}^{(2)}$ symmetry. We do this by first performing a constant gauge transformation
\be
a \rightarrow e^{-\Lambda L_0}\,a\,e^{\Lambda L_0}\,,
\qquad \bar a \rightarrow e^{\Lambda L_0}\,\bar a\,e^{-\Lambda L_0}\,,\qquad e^\Lambda\,\equiv\,\lambda\,=\,\frac{1}{2\sqrt\mu}\,.
\label{gtr}
\ee
This has the effect of changing the coefficient of $W_2$ in eq.\eqref{connections} to $\tfrac{1}{4}$, so that the coefficient of $\hat L_1\,\equiv\,\tfrac{1}{4}W_2$ is set to unity, whilst that of $L_1$ is $\lambda$, a relevant dimension-$\tfrac{1}{2}$ coupling.
 We then write the general, highest weight form of the connections,
\bea
&& a\,=\, \left(\lambda\,L_1 +\lambda_{-1}\,L_{-1}-\tfrac{1}{4}\,\lambda_{-2}\,W_{-2}\right)\,d\bar z\label{connection2}\\\nonumber\\\nonumber
&&\hspace{0.8in}+\,\left(\tfrac{1}{4} W_2 + w_1\,W_1 + \frac{3}{4 k}\,q\,W_0+ w_{-1}\,W_{-1}+\frac{w_{-2}}{4k}\, W_{-2}+\frac{\pi}{2k}\,{\cal G}\,L_{-1}\right)\,d z\,,\\\nonumber\\\nonumber
&&\bar a\,=\,- \left(-\tfrac{1}{4}\, W_{-2} + \bar w_{-1}\,W_{-1} +\frac{3}{4 k}\,\bar{q}\,W_0+ \bar w_{1}\,W_{1}-\frac{\bar w_{2}}{4\,k}\, W_{2}+
\frac{\pi}{2k}\,\bar{\cal G}\,L_{1} \right)\,d\bar z\\\nonumber\\\nonumber
&&\hspace{3.3in}- \left(\bar\lambda\,L_{-1} +\bar\lambda_1\,L_{1}+\,\tfrac{1}{4}
\bar\lambda_2\,W_{2}\right)\,dz\,.
\eea
Note that we have essentially swapped $a_z$ and $a_{\bar z}$ in eq.\eqref{connections}, so that we can focus below on the Ward identities for the holomorphic (instead of anti-holomorphic) currents.

Clearly, the connections can be easily re-expressed in terms of the ${\cal W}_3^{(2)}$ generators using \eqref{redef}. 
In accordance with the interpretation of \cite{gutkraus}, a classical solution of the form \eqref{connection2} represents a deformation of the 
${\cal W}_3^{(2)}$ CFT by the currents $G^\pm(z)$ of dimension $\tfrac{3}{2}$. Therefore the ${\rm AdS}_3$ geometry corresponding to the UV theory results when the coupling $\lambda$,  and consequently all additional parameters $\lambda_{i}, w_i, q, \ldots$, are  dialled to zero. The ${\rm SL}(2,\mathbb R) \times {\rm SL}(2, \mathbb R)$ isometry of this ${\rm AdS}_3$ is generated by two copies of $\hat L_1, \hat L_0$ and $\hat L_{-1}$ as defined in eq.\eqref{redef}.

We expect that the Ward identities of the deformed ${\cal W}_3^{(2)}$ CFT should be reproduced by the bulk field equations, namely
\be
da+a\wedge a\,=\,0,
\ee
and similarly for the barred connection. Treating the parameters $\lambda,{q}, {\cal G}, \lambda_{i}, w_i,\ldots$ as general functions of $(z,\bar z)$ and after using some of the field equations recursively, we obtain a simplified set of conditions:
\bea
&&\partial \lambda\,=\,0\,,
\qquad\qquad \lambda_{-2}\,=\,\pi\lambda\,k^{-1}{\cal G}\,,\qquad\qquad\lambda_{-1}\,=\,\tfrac{3}{2}\lambda\,k^{-1}q\,,\qquad w_1=0\,
\label{eqsa}\\\nonumber\\
&&\partial_{\bar z}\left(\,w_{-2}+\tfrac{3}{4k}\,q^2\right)
\,=\,-\pi\lambda\,\partial_{z}\,{\cal G}\,,\qquad\qquad\partial_{\bar z} w_{-1}\,=\,k^{-1}\,\lambda\,\left(-\, w_{-2}+ \tfrac{9}{4k}\,q^2 \right) 
\,\label{eqsb}\\\nonumber\\\nonumber
&&\partial_{\bar z} \,q\,=\,-4k\lambda\,w_{-1}\,,\qquad\qquad
\partial_{\bar z}\,{\cal G}\,=\,\tfrac{3}{\pi}\lambda\,\partial_{z}\,q\,.
\eea
We will show that these conditions are equivalent to the Ward identities of the ${\cal W}_3^{(2)}$ CFT deformed by spin $\tfrac{3}{2}$ currents,
\be
\delta I \,=\, -\int d^2z\,\lambda(\bar z)\left( g_+\, G^+(z)
+ g_-\, G^{-}(z)\right)\,,\label{deformn}
\ee
where $g_+$ and $g_-$ are as yet undetermined dimensionless numbers, and we have allowed $\lambda$ to be an anti-holomorphic function in $\bar z$, consistently with the field equations, \eqref{eqsa} and \eqref{eqsb}. Recall that $\lambda$ is a dimensionful coupling and is related to $\mu$, the chemical potential for spin-3 charge, as $\lambda = \frac{1}{2\sqrt\mu}$. 

\subsubsection{Ward identities for ${\cal W}_3^{(2)}$ currents}

In the presence of the deformation \eqref{deformn}, the expectation values of the currents are no longer holomorphic.
Making use of the identity, $\partial_{\bar z}\left(\frac{1}{z}\right)\,=\,2\pi\,\delta^{2}(z,\bar z)$, and the OPE's in \eqref{opes}, at linear order in $\lambda$, we find
\bea
&&\tfrac{1}{2\pi}\,\partial_{\bar z}\,\langle T_{UV}(z)\rangle_\lambda
\,=\,\tfrac{1}{2}\,\lambda\,\partial_z\,\left\langle\,\left( g_+\,G^+(z)+ g_-\,G^-(z)\right)\,\right\rangle\,,\\\nonumber\\\nonumber
&&\tfrac{1}{2\pi}\,\partial_{\bar z}\,\langle J(z)\rangle_\lambda
\,=\, -\,\lambda\,\left\langle\,\left( g_+\,G^+(z) - g_-\,G^-(z)\right)\,\right\rangle\,,\\\nonumber\\\nonumber
&&\tfrac{1}{2\pi}\,\partial_{\bar z}\,\left\langle \,\left(\, g_+\,G^+(z) - g_-\,G^-(z)\,\right)\,\right\rangle
\,=\, 2\,g_+\,g_-\,\lambda\,\left\langle\, T_{UV}(z)+ \frac{18}{\hat c}\,J(z)^2\,\right\rangle\,.
\eea
These three Ward identities must be consistent with the four (non-algebraic) conditions \eqref{eqsb}.
Direct comparison with the field equations \eqref{eqsb} then leads to the unambiguous and precise identifications,
\bea
&& w_{-1}\,=\,\frac{\pi}{2k}\,\left\langle\,\left( g_+\,G^+(z) - g_-\,G^-(z)\right)\,\right\rangle_\lambda\,,\quad\quad 
{\cal G}\,=\,\left\langle\, g_+\,G^+(z) + g_-\,G^-(z)\,\right\rangle_\lambda\,
,\nonumber\\\nonumber\\\label{ident}
&&w_{-2}\,=\,\,\left\langle\,- T_{UV}(z) -\frac{3}{4\,k}\,J(z)^2\,\right\rangle_\lambda\,,\qquad\qquad q\,=\,\,\langle J(z)\rangle_\lambda\,,
\\\nonumber\\\nonumber
&& g_+\,g_-\,=\,\pm \,\frac{1}{2\pi^2}\,.
\eea
We have used the fact that the central charges $\hat c$ and $c$ of the UV and IR  CFTs, respectively, are related as
\footnote{Here we recall that we follow the conventions of \cite{Castro:2011iw} where $c=24 k$ for the ${\cal W}_3$ CFT.}
\be
\hat c\,=\,\frac{c}{4}\,=\,6k\,.
\ee
The differing central charges can be traced to the normalizations  ${\rm Tr}\,\hat L_0^2\,=\,\tfrac{1}{2}$ and ${\rm Tr}\, L_0^2\,=\,2$ in the UV and IR field theories, respectively. Note also this means that the central charge {\em increases} along the flow, in apparent violation of the $c$-theorem in two dimensions. In \cite{gutkraus}, this puzzling phenomenon was ascribed to the presence of Lorentz violation along the renormalization group flow.

Given that $\hat L_{-1}\,=\,-\tfrac{1}{4}\,W_{-2}$, its coefficient, namely $w_{-2}$ in eq.\eqref{connection2}, should naturally be related to the energy of the configuration. This interpretation is supported by the shift of the stress tensor by $\langle J(z)^2\rangle $ in the formula for $w_{-2}$ in eq.\eqref{ident}. Precisely such a shift is encountered in the case of the BTZ black hole carrying a $U(1)$ charge \cite{krlarsen, review}. In fact, we can also match the coefficient of the shift to the level of the  current algebra generated by $J(z)$:
\be
k_{J}\,=\,\frac{4 k}{3}\,.
\ee
Here the putative black hole background  also carries a  charge associated to $J(z)$, but this charge appears to be  induced via the relevant deformation by spin-$\tfrac{3}{2}$ currents.

Another interesting feature of the analysis presented above is that the dimensionless couplings $g_\pm$ are not completely determined by this matching.
In fact, there appears to be a one parameter family of UV deformations, with $\,g_+ g_-\,=\,\pm 1/(2\pi^2)$, generating the RG flow to the IR ${\cal W}_3$ CFT.

\subsection{Thermodynamics around the UV fixed point}
 
 An important aspect of the flow from the ${\cal W}_3^{(2)}$ CFT to the IR fixed point is that the stress tensor of the IR theory is not related to the UV stress tensor. The latter acquires dimension 4 in the IR theory. The stress tensor $T_{IR}$ and spin-3 current ${\cal W}$ of the ${\cal W}_3$ CFT are actually related to the spin-1 current $J$ and spin-$\tfrac{3}{2}$ currents $G^\pm$ respectively, of the UV  CFT. Since the stress tensors of the two fixed point theories differ, thermodynamic quantities of the black hole solutions will depend on which of the two theories is used as the reference point.

Our analysis of the thermodynamics of the ${\cal W}_3$ CFT revealed multiple branches of black hole solutions. At least for small $\mu\,T$, these are classical saddle points describing the thermal properties of the ${\cal W}_3$ CFT with a chemical potential for spin-3 charge. Below we will attempt to identify the same saddle points and their thermodynamics from the UV perspective i.e. from the viewpoint of the ${\cal W}_3^{(2)}$ CFT, which should be the appropriate description when $\mu\,T\gg 1$.

The starting point of this analysis will be the gauge connection 
\eqref{connections} transformed according to eq.\eqref{gtr}. For completeness we rewrite this in the language of the ${\cal W}_3^{(2)}$ generators
\bea
&& a\,=\, \lambda\,\left[\,\sqrt 2\,(\,G^-_{1/2}-G^+_{1/2}\,)\, +\,\frac{3}{\sqrt2\, k}\,q\,(\,G^+_{-1/2}\,+\,G^-_{-1/2}\,)\,+\,\frac{\pi}{k}\,{\cal G}\,\hat L_{-1}\,\right]\,d\bar z\label{connection3}\\\nonumber\\\nonumber
&&\hspace{1.3in}+\,\left[\,\hat L_1 \,+\,\frac{3}{2\,k}\,q\,J_0 - \,w_{-2}\, \hat L_{-1}\,+\,\frac{\pi}{\sqrt 2\, k}\,{\cal G}\,(G^+_{-1/2}+G^-_{-1/2})\,\right]\,d z\,,
\eea
and similarly for the barred connection. The coefficients are naturally related to the charges of the IR CFT,
\be
q\,=\,-\frac{4\pi}{3}\,{\cal L}\,\mu\,,
\qquad w_{-2}\,=\, \frac{9}{4k}\,q^2\,,\qquad {\cal G}\,=\,4{\cal W}\,\mu^{3/2}\,,\qquad \lambda\,=\,\frac{1}{2\sqrt \mu}\,.
\label{uvir}
\ee
This demonstrates that the spin-1 charge $q$ in the UV theory is mapped to the IR stress tensor, and that the spin-$\tfrac{3}{2}$ charge in the ${\cal W}_3^{(2)}$ CFT is related to the spin-3 charge of the IR CFT. The above redefinitions can be used to rewrite the holonomy conditions \eqref{conditions} in terms of the UV variables. The two holonomy equations now take the form,
\bea
&& \frac{4}{3}\pi^2 k\,T^2\,=\,k^{-1}q^2\,-\lambda\,\pi{\cal G} - 2\lambda^2\,q\,,
\label{holagain}\\\nonumber\\\nonumber
&& \pi^2{\cal G}^2 + 4k^{-1}\,q^3 + 6\pi\,\lambda\, {\cal G}\,q
+24\,\lambda^2\, q^2 - 4\pi k\,\lambda^3\,{\cal G}\,=\,0\,,
\eea
where we have used the Chern-Simons equations of motion to set $w_{-2}= 9q^2/4k$. It is now easy to see what happens in $\lambda \to 0$ limit, which we would like to identify as the UV limit, where the relevant deformation of the ${\cal W}_3^{(2)}$ CFT disappears
\footnote{Strictly speaking, since $\lambda$ is a dimensionful parameter, 
the $\lambda\to 0$ limit should be interpreted as taking $\lambda/\sqrt{T} \ll 1$.}. In this limit the holonomy conditions simplify and become $\lambda$-independent, yielding
\be
q^2\,\big|_{\lambda\to 0}\,\to \,\frac{4}{3}\pi^2k\,T^2\,,\qquad\qquad\qquad {\cal G}\big|_{\lambda\to 0}\,=\,\pm\frac{2k}{\pi}\left(\frac{2}{\sqrt 3}\,\pi\,T\right)^{3/2}\,.
\ee
Notice that although there are four roots for the system in this limit, only two are real (these are branches III and IV) and we need to choose the negative sign for ${\cal G}$ to pick out branch-III that has the lower free energy. This result is significant in several ways: 
\\\\
(i) Firstly, the theory has a VEV for the spin-$\tfrac{3}{2}$ currents even when $\lambda \to 0$ and this expectation value is consistent with dimensional analysis at high temperature. 
\\\\
(ii) Secondly, the fact that the expectation value ${\cal G} \neq 0$ generically (for any $\lambda$) indicates that the symmetry generated by $J(z)$ is spontaneously broken, even at arbitrarily high temperature. 
This is because ${\cal G} \sim \langle g_+\,G^+\rangle$ and the operator $G^+$ carries charge $+1$ under $J$. 

It is worth bearing in mind that at the level of the Chern-Simons equations of motion ${\cal G}$ is not required to be non-zero  (see \eqref{eqsa} and \eqref{eqsb} taking the parameters to be independent of $(z,\bar z)$). The requirement that ${\cal G} \neq 0$ even in the $\lambda \to 0$ limit is actually imposed by the holonomy conditions which we now recognize as integrability conditions resulting in consistent thermodynamics. The condensate ${\cal G}$ is allowed to vanish only when $ T = 6 \lambda^2/\pi$, which is, of course, the temperature at which the spin-3 charge 
 also vanished for branch III of black hole solutions.
\\\\
(iii) Finally, we have learnt from the Ward identities (eq.\eqref{ident}) of the ${\cal W}_3^{(2)}$ CFT that the energy of the state, as measured near the UV fixed point of the flow, is given by
\be
2\pi\hat{\cal L}\,\equiv\,-\,\langle T_{UV}\rangle\,=\,\left(w_{-2}+
\frac{3}{4k}\,q^2\right)\,=\,\frac{3}{k}\,q^2\,\big|_{\lambda\to 0}\,\to \,4\pi^2k T^2,
\ee
and this now has the correct high temperature behaviour for a 2d CFT, in contrast to the result \eqref{highT} for branch III from the perspective of the IR CFT. In particular, this indicates that the so-called branch III of black hole solutions is physical and that the non-trivial phase diagram found previously should have a natural explanation from the viewpoint of the finite temperature RG flow between the two CFT's in question.

\paragraph{\underline {Thermodynamics}:} Repeating the steps outlined in Section\eqref{thermoaction} we can try to obtain all thermodynamical quantities in the deformation of the UV CFT. At the outset, we might expect that the on-shell actions in the two descriptions should be the same. However, we recall that in flowing from the IR to the UV conformal fixed point, we performed a switch $a_z \leftrightarrow 
a_{\bar z}$ (see eq.\eqref{connection2}) which results in a sign change. Then using the boundary action \eqref{boundary} the {\em on-shell} action in terms of the UV variables is
\be
\hat I_{\rm on-shell}\,=\, -\,\beta\,\left(2\pi\hat{\cal L}  + \,6\lambda^2\,q\right).
\ee
We note in passing that this on-shell action\footnote{Curiously, the on-shell action can also be written simply as $\hat I_{\rm on-shell}\,=\,\beta(2\pi {\cal L}-2\pi \hat{\cal L})$.}
 could have been interpreted as the sum of two terms:  An energy density given by $2\pi\hat{\cal L}$ and a chemical potential $\sim\lambda^2$ for the spin-1 charge $q$. The significance of this is unclear, since the on-shell action cannot be used to vary with respect to the potentials to compute thermodynamical quantities. Therefore, as before, we follow the procedure of constructing the general variation of the Chern-Simons action. This first requires a Legendre transform of the on-shell action, so that the independent thermodynamic potentials are $\beta$ and $\lambda$,
\be
\hat I = \hat I_{\rm on-shell} - \pi{\cal G}\,\lambda\beta\,.
\ee
Interestingly we find once again that the action has to be defined with a shift $\hat I _{\rm th}\,=\,\hat I -4\pi^2 k T^2$ in order to obtain the correct value of the energy upon varying with respect to $\beta$. Subsequent to these two steps, a general variation of the thermodynamical action yields
\bea
&&\frac{i}{2\pi}\,\delta \hat I_{\rm th}\,=\,\,4\pi{\cal G}\,d\hat\alpha
\,+ \,4\pi\hat {\cal L}\,d\tau\,,\\\nonumber\\\nonumber
&&\hat\alpha\,\equiv\,-\lambda\,\tau\,.
\eea
Now we can interpret the high temperature phase of the black hole branch III as the ${\cal W}_3^{(2)}$ CFT with a chemical potential $\lambda$ for the spin-$\tfrac{3}{2}$ currents at finite temperature. 
The value of the grand potential for the UV theory is 
\be
\hat \Phi\,=\, - 4\pi\hat{\cal L} \,+\,2\pi{\cal G}\,\lambda\,.
\ee
In fact, using the relations \eqref{uvir} between the UV and IR variables and the holonomy condition ${\rm Tr}(\beta^2\,a_t^2)\,=\,-8\pi^2$, we find a remarkable identity satisfied by the grand potentials for the UV and IR CFT's (all hatted quantities refer to the UV fixed point theory):
\be
\boxed{\,\,\Phi + \hat\Phi\,=\, -8\pi^2k\,T^2\,\,}\,.
\ee 
This equation applies at the level of the thermodynamic variables, upon evaluating the quantities on any given solution or saddle point at fixed $\mu$ and $T$. Similarly, the entropy that follows from this analysis, $\hat S\,=\, (8\pi\hat{\cal L}-6\pi\lambda{\cal G})/T$, is also related to the IR definition of entropy via,
\be
\boxed{\,\,S + \hat S\,=\, 16\pi^2k\,T^2\,\,}\,.
\ee
The above identities work to ensure that a sensible thermodynamic interpretation exists for the dominant branches at small and large $\mu T$ in terms of either the IR or UV CFT. For example, when $\mu T\gg 1$, it may be explicitly verified that $\Phi$ evaluated on branch-III scales as $T^{3/2}$ while $\hat \Phi \approx -8\pi^2T^2k$. In the same way, the entropy $\hat S$ displays the expected behaviour at high temperature,
\bea 
&&\hat S\,=\,8\pi\sqrt{2\pi\hat{\cal L}\,\,k}\,\,\hat f( y)\qquad\longrightarrow\qquad 16\pi^2 T\,k\left(1+3^{1/4}\frac{\lambda}{\sqrt {2\pi T}} +\ldots\,\right)
\\\nonumber
&&y\,\equiv\, \tfrac{3}{8}\sqrt{\tfrac{3\pi}{2k}}\,\,\frac{{\cal G}^2}{{\hat {\cal L}}^{3/2}}\,,\\\nonumber\\\nonumber
&&9\,y\,(2- y)\, {\hat f}^{\,\prime\, 2}\,=\,1-\hat f^{\,2}\,,\qquad\qquad\hat f(1) =1\,.
\eea
The dimensionless variable $y$ is identical to the one appearing in eq.\eqref{yentropy}, but expressed in terms of the UV charges. 
To obtain the differential equation determining the function $\hat f(y)$  in the above equation
we have used the following thermodynamic relations
\begin{equation}
\tau = \frac{i}{8\pi^2} \frac{\partial \hat S}{\partial \hat {\cal L }}, \qquad
\hat \alpha = \frac{i}{8\pi^2} \frac{\partial\hat S}{\partial {\cal G}} ,
\end{equation}
together with the first holonomy condition given in \eqref{holagain}. 
The limit $\lambda \to 0$ corresponds to $y\to 1$, whilst in the low temperature limit, $y=2$. Finally we present the high temperature expansion $(\mu T\gg1)$ of the free energy
of branch-III, and for comparison we also quote the result for the BTZ-branch that dominates when $\mu T\ll 1$:\\\\
{\underline{BTZ-branch: $\mu T\ll 1$}}
\bea
\ln\,Z\,=\,\frac{4i\pi k}{\tau}\,\left[1-\frac{4}{3}\,\frac{\alpha^2}{\tau^4}\,+\,\frac{160}{27}\,\frac{\alpha^4}{\tau^8}
-\frac{1088}{27}\,\frac{\alpha^6}{\tau^{12}}+\ldots\right]\,,
\eea
{\underline{Branch-III: $\mu T\gg 1$}}
\bea
&&\ln\,Z\,=\,\frac{4i\pi k}{\tau}\,\left[1+ 
\tfrac{4}{3^{3/4}}\,z
+\,2\sqrt 3\,z^2 \,+ \,3^{3/4}\,z^3+3\,z^4 +\tfrac{3^{5/4}}{8}\,z^5+\,\tfrac{3\sqrt 3}{4}\,z^6+\ldots\right]\nonumber\\\\\nonumber
&& z\,\equiv\,\left(\frac{-i \hat\alpha^2}{\tau}\right)^{1/2}\,=\,
(8\pi\mu T)^{-1/2}\,.\label{uvexpansion}
\eea
The first few orders in the expansion of the partition function of the BTZ-branch have already been matched with a perturbative expansion (in $\mu$) about the 
IR CFT with ${\cal W}_3$ symmetry \cite{Gaberdiel:2012yb}. Whether a similar expansion (for large $\mu$) can be set up for the deformation of the ${\cal W}_3^{(2)}$ CFT in the UV and matched to \eqref{uvexpansion}, remains an open question.
\section{Discussion}
Our study of the black hole branches carrying spin-3 charge has highlighted several interesting aspects of this system and also raised certain puzzles.
The most interesting question, of course, is what features of the picture we have obtained above will generalize to ${\rm SL}(N, \mathbb R)$ Chern-Simons theories. It is clear  that for $N \ge 3$, the holonomy conditions become progressively more complicated \cite{Kraus:2011ds, bct}, and hence the multiplicity of solutions increases. This means that the phase  structure for solutions carrying higher spin charges will be intricate, and contributions from conical defect-like  states may also become relevant \cite{Castro:2011iw}. Furthermore, a chemical potential for a higher spin charge will generically correspond to an irrelevant deformation and will modify the asymptotics in much the same way as in the spin-3 case. 

It is particularly striking that the Chern-Simons connection describing the spin-3 deformation can be viewed from the perspective of either a UV or IR  CFT, and describes black holes within an RG flow connecting two CFT's. The fact that a description from the viewpoint of either CFT can be obtained by simply exchanging the terms that correspond to `background' and `deformation', providing a simple map between the variables of the two fixed points CFT's, is reminiscent of Legendre transform pairs of field theories within the general context of AdS/CFT \cite{Klebanov:2002ja,Witten:2001ua,Klebanov:1999tb}. It would be interesting if this could be made precise. 

We have also presented the high temperature expansion of the free energy of the branch-III black hole solutions. A direct verification of this expansion using the properties of the ${\cal W}_3^{(2)}$ CFT would help to validate the picture we have presented. It would also confirm that the duality between the gravitational (higher spin) theory and the deformed ${\cal W}_3$ CFT extends to finite $\mu$ (and would not only be valid for small $\mu$ alone). Such a computation could be performed along the lines of \cite{Kraus:2011ds,Gaberdiel:2012yb}, perhaps first using the free field representation of the ${\cal W}_3^{(2)}$ algebra as a possible starting point \cite{deBoer:1993iz}. This would help to
shed more light on the nature of the gravitational theory that is dual to the UV fixed point theory which has bosonic spin-$\tfrac{3}{2}$ currents. The relation between these currents and those of a bulk higher spin theory remains mysterious.

\acknowledgments 
We would like to thank Chethan Krishnan, Gautam Mandal, Shiraz Minwalla and Sandip Trivedi for comments and enjoyable discussions. 
SPK would like to thank the Centre for High Energy Physics, Indian Institute of Science, Bangalore, and the Dept. of Theoretical Physics at the Tata Institute of Fundamental Research, Mumbai, for their hospitality and providing a stimulating atmosphere for carrying out part of this work. SPK and MF were supported in part by STFC grant ST/G000506/1.

\startappendix
\Appendix{Black hole gauge metric parameters}
\label{appb}
The metric for the spin-3 charged solution written in black hole gauge 
\eqref{bhgauge} is written in terms of the parameters $a_\phi, b_\phi, a_t$ and $b_t$ where,
\bea
&& a_t\,=\,(C-1)^2\,\left(4\gamma -\sqrt C\right)^2\,,\\\nonumber
&& a_\phi\,=\,(C-1)^2\,\left(4\gamma +\sqrt C \right)^2\,,
\\\nonumber
&& b_t\,=\,16\gamma^2(C-2)(C^2-2C+2) - 8\gamma\sqrt{C}\,(2C^2-6C+5)+
C(3C-4)\,,\\\nonumber
&& b_\phi\,=\,16\gamma^2(C-2)(C^2-2 C+2) + 8\gamma\sqrt{C} (2C^2-6 C+5)+ C(3C-4)\,.
\eea

\Appendix{The ${\cal W}_3^{(2)}$ algebras}
\label{appa}
We summarise the basic properties of the ${\cal W}_3^{(2)}$ algebra or the Polyakov-Bershadsky algebra \cite{Polyakov:1989dm, Bershadsky:1990bg} relevant for us.  The OPE of the associated currents for the UV conformal field theory was inferred in \cite{ammon} from the action of gauge transformations on the Chern-Simons connections. The algebra consists of the stress tensor $T(z)$, two {\em bosonic} spin-$\tfrac{3}{2}$ currents $G^\pm(z)$ and one spin-1 current $J(z)$:
\bea
&& T(z)T(0)\sim \frac{\hat c}{2\,z^4}+\frac{2}{z^2}\,T(0)+\frac{1}{z}\partial T(0)\,,\qquad\qquad J(z)\,J(0) \sim -\frac{\hat c}{9\,z^2}\,,\label{opes}\\\nonumber
&& T(z)J(0)\sim \frac{1}{z^2}J(0)+\frac{1}{z}\partial J(0)\,,\qquad\qquad
T(z)G^\pm(0)\,\sim\, \frac{3}{2\,z^2}G^\pm(0)+\frac{1}{z}\partial G^\pm(0)\,,\\\nonumber
&&J(z)G^\pm(0)\,\sim \,\pm \frac{1}{z}G^\pm(0)\,,\\\nonumber
&& G^+(z)G^-(0)\sim -\frac{\hat c}{3\,z^3}+\frac{3}{z^2}\,J(0) -\frac{1}{z}\left( T(0) -\frac{3}{2}\partial J(0) +\frac{18}{\hat c}\,J(0)^2\right)\,,
\eea
where $\hat c = c/4$, and $c$ is the central charge of the ${\cal W}_3$ (IR) CFT. The non-linear term in the $G^+ G^-$ OPE vanishes in the large $\hat c$ limit (which is the limit in which the dual classical (higher spin) gravity description applies). From the OPE's we can easily deduce commutation relations for the modes in the expansions
\be
T(z)\,=\,\sum L_n\,z^{-n-2}\,,\qquad J(z)\,=\,\sum J_n\,z^{-n-1}\,,\qquad G^{\pm}(z)\,=\,\sum G^{\pm}_{r}(z)\,z^{-r-\tfrac{3}{2}}\,.
\ee
It is important note that in principle the mode expansion for the spin-$\tfrac{3}{2}$ currents can have integer (Ramond) or half-integer (Neveu-Schwarz) moding. We find it necessary to pick the half-integer moding, in order to avoid having  central terms in the global part of the algebra and so that this global part is isomorphic to the ${\rm SL}(3, \mathbb R)$ algebra. The important commutation relations are:
\bea
&&[J_n, G_m^\pm]\,=\,\pm G^{\pm}_{n+m}\,,\qquad[J_n,J_m]\,=\,-\frac{\hat c}{9}\,n\,\delta_{n+m,0}\,,
\label{w32}\\\nonumber
&&[\hat L_n, G^\pm_m]\,=\,\left(\tfrac{1}{2}n-m\right)\,G^\pm_{n+m}\,,\qquad [\hat L_n,J_m]\,=\,-m\,J_{n+m}\,,
\\\nonumber
&&[G_n^+,G^-_m]=-\frac{\hat c}{6}\left(n^2-\tfrac{1}{4}\right)\delta_{n+m,0} - \hat L_{n+m}+\tfrac{3}{2}(n-m)J_{n+m}\,,
\eea
For comparison we also write down the commutation relations obeyed by the global part of the ${\cal W}_3$ algebra {i.e.} ${\rm SL}(3, \mathbb R)$:
\bea
&&[L_i,L_j]\,=\,(i-j)L_{i+j}\,,\qquad[L_i,W_m]\,=\,(2i-m)W_{i+m}\,,\\\nonumber
&&[W_m, W_n]\,=\,-\frac{1}{3}(m-n)(2 m^2 + 2 n^2 - mn-8) L_{m+n}\,.
\eea

\end{document}